\documentclass[conference]{IEEEtran}
\IEEEoverridecommandlockouts

\usepackage{cite}
\usepackage{amsmath,amssymb,amsfonts}
\usepackage{algorithmic}
\usepackage{graphicx}
\usepackage{textcomp}
\usepackage{xcolor}
\usepackage{booktabs}
\def\BibTeX{{\rm B\kern-.05em{\sc i\kern-.025em b}\kern-.08em
    T\kern-.1667em\lower.7ex\hbox{E}\kern-.125emX}}

\usepackage{fancyhdr}
\begin{document}

\title{Augmenting Interviewer Judgments of Patient Experience with Automatic Language Analysis\\
}

\author{
\IEEEauthorblockN{
Aowen Shi\textsuperscript{1},
Michal Balazia\textsuperscript{1},
Danilo Postin\textsuperscript{2},
René Hurlemann\textsuperscript{2}\\
Jan Alexandersson\textsuperscript{3},
Fran\c{c}ois Br\'emond\textsuperscript{1},
Philipp Müller\textsuperscript{4}
}

\IEEEauthorblockA{
\textsuperscript{1}\textit{INRIA Université Côte d'Azur}, Valbonne, France, aowen.shi@inria.fr, michal.balazia@inria.fr, francois.bremond@inria.fr
}

\IEEEauthorblockA{
\textsuperscript{2}\textit{Carl von Ossietzky University of Oldenburg}, Germany, danilo.postin1@uni-oldenburg.de, rene.hurlemann@uni-oldenburg.de
}

\IEEEauthorblockA{
\textsuperscript{3}\textit{German Research Center for Artificial Intelligence}, Saarbrücken, Germany, jan.alexandersson@dfki.de
}

\IEEEauthorblockA{
\textsuperscript{4}\textit{Max Planck Institute for Intelligent Systems}, Stuttgart, Germany, phmueller@is.mpg.de
}
}

\maketitle
\thispagestyle{fancy}

\begin{abstract}
Understanding how psychiatric patients subjectively experienced a clinical conversation is important for feedback and alliance-related process monitoring. While interviewers form post-session judgments about patient experience, these judgments do not always match patients' self-reports. Automatic approaches for predicting perceived interaction quality from conversation have been proposed, but it remains unclear whether such approaches can complement human judgment rather than simply replicate it. To address this gap, we evaluate a clinician-support framework in which post-session interviewer ratings are combined with automatic language-based predictions to estimate patient-reported interaction quality in free clinical interviews. We assess this integration across multiple standard model types, including Ridge, SVR, MLP, GRU, and BiLSTM, all trained on sentence embeddings extracted from dyadic transcripts of 107 free conversations between psychiatric patients and interviewers. Our results show that combining interviewer judgments with model predictions through simple averaging yields the strongest overall performance. The interviewer-only baseline reached a Pearson correlation of 0.365. Among fully automatic models, Ridge achieved the strongest Pearson correlation (\(r\) = 0.286), while BiLSTM achieved \(r\) = 0.270. The strongest result was obtained by BiLSTM interviewer integration (\(r\) = 0.403). Our findings suggest that automatic language analysis and interviewer judgment capture complementary aspects of patient experience and that their combination provides a more accurate approximation of the patient's own report than either source alone. 
\end{abstract}


\section{Introduction}

How patients feel during interactions with clinic personnel is crucial for patient satisfaction and therapy outcomes.
This holds true across clinical disciplines~\cite{stewart1995effective, fuertes2007physician, priebe2008therapeutic}. In psychotherapy, the quality of interaction between patient and therapist is conceptualized as the \textit{therapeutic alliance}~\cite{bordin1979generalizability, horvath2011alliance}. The quality of therapeutic alliance was shown to be a consistent predictor of therapy outcome across all major schools of psychotherapy~\cite{fluckiger2018alliance, fluckiger2020assessing}. Due to its importance, it is crucial for clinic personnel to assess the quality of a patient's experience. Despite this importance, several studies have shown that it can be difficult for clinical personnel to accurately judge the quality of a patient's experience from interacting with them, including underestimating perceived alliance and overestimating positive patient emotions~\cite{hartmann2015accuracy, atzil2019therapists}.

Affective computing holds great promise to provide automatic approaches that can assist in understanding patient experience.
Indeed, approaches to estimate conversation quality have been proposed in a variety of scenarios, including group discussions \cite{muller2018detecting}, speed dating interactions~\cite{vargas2021individual}, and in mental health conversations such as counseling and psychotherapy~\cite{Goldberg2020, Lin2025, ryu2023natural}. 
With the success of transformer-based language models, text-based approaches have seen increasing success~\cite{devlin2019bert, reimers2019sentence}. 
At the same time, these methods offer advantages with respect to patient comfort and subjective privacy compared to video-based approaches~\cite{torous2017ethical}.
However, these approaches for estimating conversation quality were only investigated separately from human judgments.
As a result, it remains unclear to what extent such approaches might be able to help interviewers in improving their understanding of patient experience.

In our work, we close this gap by investigating whether automatic language-based approaches to interaction quality estimation can be successfully combined with judgments obtained from the patient's interviewer.
We systematically evaluate this integration across five standard model types, all operating on sentence embeddings extracted separately for patient and interviewer streams. The resulting predictions are subsequently averaged with the interviewer's judgment of patient experience.

We evaluate this framework on 107 free clinical interviews from the German cohort of the MePheSTO corpus~\cite{konig2022multimodal}. Our results show that interviewer integration (Pearson correlation $r=0.403 \pm 0.030$) consistently outperforms both the interviewer-only baseline ($r=0.365$) and the best fully automatic model ($r=0.286 \pm 0.044$), demonstrating that the two sources carry complementary. Additional speaker-stream analyses show that the performance of the fully automatic approach is primarily driven by the interviewer-side transcript.

\section{Related Work}
\subsection{Patient-experienced interaction quality and clinician}

Many psychotherapy studies have demonstrated the significant clinical importance of how patients experience therapeutic interactions. In particular, the therapeutic alliance is one of the process variables most closely associated with treatment outcomes across psychotherapy models, making patients' experience of the therapeutic relationship crucial not only at the descriptive level but also at the prognostic level~\cite{fluckiger2018alliance, fluckiger2020assessing}. Research on the breakdown, repair, and feedback mechanisms of the therapeutic alliance further confirms that monitoring patients' experience of the interaction is crucial for retaining patients, adjusting treatment plans, and improving treatment outcomes~\cite{safran2000resolving, Eubanks2018, Lambert2018ROM}.

At the same time, research on the psychotherapy process has shown that patient and therapist perspectives on the patient's subjective experience and the quality of the therapeutic relationship are related but not interchangeable. 
For example, Hartmann et al.~\cite{hartmann2015accuracy} found that therapists tend to underestimate patients' perceived sense of therapeutic alliance, while Atzil-Slonim et al.\ showed that therapists tracked patients' emotions with some accuracy but still tended to overestimate positive emotions~\cite{atzil2019therapists}. 
Research on patient-therapist agreement and disagreement shows that stronger alliance agreement is frequently linked to better symptom trajectories, reduced dropout rates, or better treatment outcomes~\cite{Jennissen2020, MosheCohen2024}. 
Overall, this implies that (1) clinician judgments of the interaction are informative but may misrepresent the patient's subjective experience, and (2) achieving a higher agreement between clinician judgments and patient subjective experience has the potential to improve treatment. 
Our study builds on these observations and explores whether automatic language analysis has the potential to improve clinician judgments of subjective patient experience.

\subsection{Prediction of interaction quality from conversation}

The development of automatic approaches for predicting interaction quality has received growing interest in affective computing. 
Previous research has modeled constructs such as rapport, engagement, and perceived conversation quality from behavioral evidence in natural interactions from a variety of behavior modalities~\cite{muller2018detecting, raman2023perceived, vargas2021individual}.

In the field of psychotherapy, research indicates that spoken interaction contains measurable process signals. Earlier studies used natural language processing techniques to quantify the coding of psychotherapy processes and the assessment of therapist skills. For example, in motivational interviewing, these studies demonstrated that clinically meaningful session qualities can be inferred from lexical content extracted from audio recordings~\cite{Imel2015, imel2019design, tanana2016comparison}. Recent work has also explored AI-generated patient simulations for assessing motivational interviewing sessions, further illustrating the growing role of language-based AI methods in psychotherapy process assessment~\cite{yosef2024assessing}. More directly relevant to our research context, Goldberg et al.~\cite{Goldberg2020} showed that machine learning models trained on session linguistic content could predict client-rated therapeutic alliance from psychotherapy recordings, while later work further identified alliance-related language markers in transcripts and developed language-model-based frameworks for inferring working alliance from psychotherapy dialogue~\cite{Lin2025, ryu2023natural}. These studies share our general framing of predicting patient-reported relational experience from conversational language, but treat automatic models as standalone systems evaluated against a held-out patient label. A key question they leave open is whether such automatic estimates carry information that is independent of, and therefore complementary to, the judgment of a human observer who was present during the interaction. This is the gap our study directly addresses.

Recent multimodal studies suggest that incorporating audio and video in certain contexts may further improve the predictive performance of therapeutic alliance~\cite{aafjes2025predicting}.
However, such approaches introduce additional technical and privacy constraints that limit their applicability in routine clinical settings. Our work therefore focuses on language-based modeling, which offers a more scalable and privacy-preserving starting point for automatic interaction quality analysis.

\section{Dataset and Labels}
\subsection{Clinical Interviews from the MePheSTO corpus}

The present analysis is based on the German cohort of the MePheSTO corpus~\cite{konig2022multimodal}.
The German portion consists of audio and video recordings of \textit{107} dyadic clinical sessions between psychiatric patients and interviewers. The patient sample consisted of individuals screened with the structured clinical interview for DSM-5 (SCID-5) and meeting criteria for a Major Depressive Episode. Patients were recruited from inpatient services of the Karl-Jaspers-Klinik (Bad Zwischenahn, Germany), in cooperation with the Department of Psychiatry, University of Oldenburg. The interviews were free-format and unstructured: participants could talk about any topic of their choice, and natural, unconstrained speech was recorded. The interviews lasted \textit{47} minutes on average, with durations ranging from \textit{16} to \textit{81} minutes. All sessions were conducted by trained research assistants (Medicine and Psychology graduate students) in a video-mediated setup in which the patient and interviewer were located in separate rooms and interacted remotely. The setup yields separate patient-side and interviewer-side audio recordings rather than a single mixed-channel recording, which enables the speaker-stream comparisons reported in this paper.

The patient sample comprises \textit{38} patients (\textit{15} female, \textit{23} male) with a mean age of \textit{33.84} years (SD = \textit{13.39}, range = \textit{18}--\textit{62}). Baseline depressive symptom severity, assessed with the BDI-II, had a mean of \textit{30.63} (SD = \textit{11.92}). 
Across patients, the number of interview sessions ranged from \textit{1} to \textit{4}, with a mean of \textit{2.82} interviews per patient.
After each interview session, patients and interviewers completed a post-interaction questionnaire to evaluate the interaction. Responses were recorded on visual analog scales (VAS) ranging from 0 to 100. The present study uses three of these post-interaction items as the label of clinical experience.

\subsection{Interaction quality measure}

We operationalize free interview experience using three post-interaction items from the MePheSTO protocol~\cite{konig2022multimodal}. The MePheSTO post-interaction questionnaire comprises 10 items and draws on the Working Alliance Inventory (WAI)~\cite{horvath1989development} as a conceptual basis for capturing alliance-related aspects within a study design involving up to four free-format, unstructured interviews. The subset analyzed in this study was restricted to the only three items available in parallel patient-report and interviewer-estimate form, allowing us to compare patients' self-reported experience with interviewers' perspective-taking judgments of the same interaction. In addition, these items capture complementary aspects of a supportive interview experience: perceived helpfulness of the conversation, ease of sharing personal information, and perceived mood change after the interaction. Table~\ref{tab:items} summarizes the original item wording and the interpretive notes used in this study.
\begin{table*}[!t]
    \centering
    \caption{Post-interaction questionnaire items analyzed in this study.}
    \label{tab:items}
    \scriptsize
    \renewcommand{\arraystretch}{1.1}
    \begin{tabular}{@{}p{0.10\textwidth}p{0.26\textwidth}p{0.26\textwidth}p{0.29\textwidth}@{}}
        \toprule
        \textbf{Target} & \textbf{Patient wording} & \textbf{Interviewer wording} & \textbf{Interpretive note} \\
        \midrule
        Mood &
        After this conversation, I feel better. &
        I have the feeling that my conversation partner feels better after our conversation. &
        Perceived mood-related impact of the conversation. \\

        Helpfulness &
        I have the feeling that what we discussed today was helpful. &
        I have the feeling that what we discussed today helped my conversation partner. &
        Perceived helpfulness or benefit of the conversation. \\

        Personal sharing &
        Today, it was easy for me to share personal information with my conversation partner. &
        I have the feeling that it was easy today for my conversation partner to share personal information with me. &
        Relational openness and enabled personal disclosure. \\
        \bottomrule
    \end{tabular}
\end{table*}

For each session $i$, the primary target is the overall patient-reported score, defined as the mean of three patient-side items
\begin{equation}
    y_i^{\mathrm{P}} = \frac{y_i^{\mathrm{mood}} + y_i^{\mathrm{helpful}} + y_i^{\mathrm{personal}}}{3}
\end{equation}
where $y_i^{\mathrm{mood}}$, $y_i^{\mathrm{helpful}}$, and $y_i^{\mathrm{personal}}$ denote the patient-side ratings of "After this conversation, I feel better,"
"I have the feeling that what we discussed today was helpful,"
and "Today, it was easy for me to share personal information with my conversation partner" respectively. Similarly, $y_i^{\mathrm{I}}$ represents the interviewer's overall evaluation, calculated as the average of three corresponding interviewer-side items. This composite is used as the main target because the three items capture complementary aspects of perceived interview quality: mood improvement, helpfulness, and relational openness. Conceptually, these aspects relate to Bordin's three dimensions of the therapeutic alliance, bond, goals, and tasks~\cite{bordin1979generalizability}. However, the composite does not fully operationalize therapeutic alliance. This is intentional because the MePheSTO questionnaire was constrained to a small number of post-interaction items and because the recorded sessions were free clinical interviews rather than ongoing psychotherapy sessions. We therefore interpret the composite as a patient-reported interaction-quality target related to therapeutic alliance, rather than as a complete scale for measuring it.
\begin{figure}[!t]
    \centering
    \includegraphics[width=252pt]{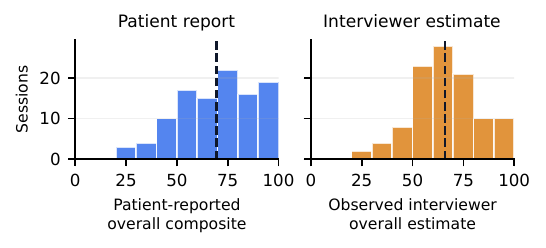}
    \caption{Distributions of the patient-reported overall composite and the observed interviewer overall estimate on the 106 sessions. Dashed lines indicate the mean.}
    \label{fig:rating_histograms}
\end{figure}

All \textit{107} sessions have complete patient-reported ratings, while one session is missing the interviewer post-interaction questionnaire, leaving \textit{106} sessions used for all analyses. Among these \textit{106} sessions, the mean patient-side composite is \textit{69.5} (SD = \textit{18.9}, range = \textit{24.3}--\textit{100.0}) and the mean interviewer-side composite is \textit{66.1} (SD = \textit{15.9}, range = \textit{27}--\textit{100.0}). Figure~\ref{fig:rating_histograms} shows the distributions of $y_i^{\mathrm{P}}$ and $y_i^{\mathrm{I}}$ across the 106 sessions. Both distributions are left-skewed, with patients reporting slightly higher scores on average than interviewers.

\section{Methods}
\begin{figure*}[!t]
    \centering
    \includegraphics[width=\textwidth]{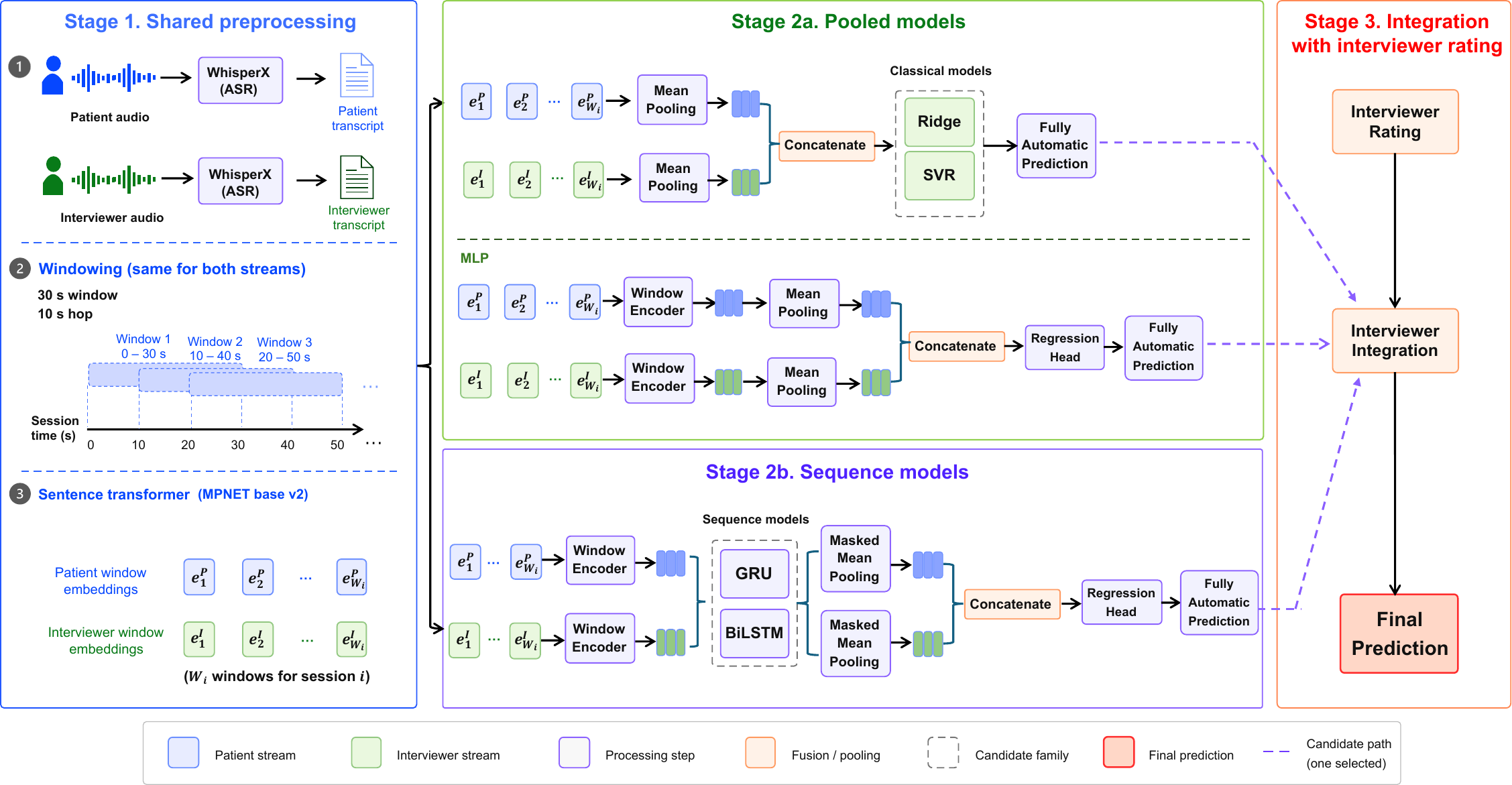}
    \caption{Overview of our approach to predict patient experience by incorporating automatic language analysis with interviewer ratings. Stage 1 consists of three sequential preprocessing substeps: transcription, windowing, and embedding generation.
    }
    \label{fig:architecture}
\end{figure*}

Figure~\ref{fig:architecture} gives an overview of our methodology. For each session, we extract separate language representations for the patient and interviewer streams, train regression models to predict the patient-reported experience, and then combine the resulting predictions with the interviewer's post-session rating by averaging. Our goal is to systematically evaluate whether the integration of interviewer ratings with automatic prediction models is able to consistently improve the accuracy of estimating patients' subjective experience.

\subsection{Preprocessing and dyadic language representation}

Separate speaker channels make role-specific transcripts available for both speakers. Patient and interviewer speech streams are transcribed with WhisperX~\cite{bain2023whisperx}. The transcript text is segmented into fixed overlapping 30-second windows with a 10-second stride. The 30-second window with a 10-second stride was used as a preprocessing choice rather than an optimized hyperparameter. Since full interviews are too long for single-input encoding, windowing converts each conversation into local temporal segments. The overlap reduces sensitivity to window boundaries. Each window is embedded using MPNet base v2, a multilingual sentence transformer~\cite{reimers2019sentence} that supports over 50 languages, including German. We selected this model because the MePheSTO corpus was recorded in German. Using a multilingual model ensures that the embeddings capture the semantic content of the sessions without requiring language-specific fine-tuning. The sentence transformer yields an aligned 768-dimensional window-level embedding sequence for the patient and interviewer transcripts. 

\subsection{Regression models}

We compare two types of regression approaches that differ in how they summarize each speaker stream: pooled models that compress the full stream into a single session-level vector and sequence models that preserve temporal structure before pooling. This distinction is motivated by the question of whether the ordering and dynamics of language within a session carry additional predictive information beyond an aggregate summary. To assess whether any benefit of integration is specific to a particular modelling choice or holds more generally, we evaluate five models: pooled models Ridge, SVR and MLP, and sequence models GRU and BiLSTM.

Given window $w$ in session $i$ of $W_i$ aligned windows, let $\mathbf{e}^{P}_{iw},\mathbf{e}^{I}_{iw} \in \mathbb{R}^{768}$ denote the patient and interviewer window embeddings, respectively, for the pooled models. Similarly, for the sequence models, we define $\mathbf{h}^{P}_{iw},\mathbf{h}^{I}_{iw} \in \mathbb{R}^{768}$ as the patient‑side and interviewer‑side encoded window‑level hidden representations obtained after the shared window encoder and the selected sequence encoder.

\textbf{Pooled models.} For Ridge and SVR, each speaker stream is mean-pooled directly from the window embeddings:
\begin{equation}
    \mathbf{z}_i^{\mathrm{pool}} =
    \left[
    \frac{1}{W_i}\sum_{w=1}^{W_i}\mathbf{e}^{P}_{iw};
    \frac{1}{W_i}\sum_{w=1}^{W_i}\mathbf{e}^{I}_{iw}
    \right]
\end{equation}

For the MLP, a shared window encoder $\phi(\cdot)$ is first applied to each window embedding, and the encoded features are then mean-pooled within each stream:
\begin{equation}
    \mathbf{z}_i^{\mathrm{mlp}} =
    \left[
    \frac{1}{W_i}\sum_{w=1}^{W_i}\phi\!\left(\mathbf{e}^{P}_{iw}\right);
    \frac{1}{W_i}\sum_{w=1}^{W_i}\phi\!\left(\mathbf{e}^{I}_{iw}\right)
    \right]
\end{equation}

\textbf{Sequence models.} For sequence models, we preserve the ordered window embeddings for each speaker; patient and interviewer sequences are encoded separately, pooled with masked mean pooling, concatenated by late fusion, and mapped to a scalar prediction with a regression head:
\begin{equation}
    \mathbf{z}_i^{\mathrm{seq}} =
    \left[
        \frac{1}{W_i}\sum_{w=1}^{W_i}\mathbf{h}^{P}_{iw};
        \frac{1}{W_i}\sum_{w=1}^{W_i}\mathbf{h}^{I}_{iw}
    \right]
\end{equation}

\subsection{Prediction settings}

We investigate three different ways to utilize interviewer intuition and automatic model predictions.

\textbf{Raw interviewer.} The interviewer rating $y_i^{I}$ is used directly as a proxy for the patient's report, serving as the human-judgment baseline. 

\textbf{Fully automatic.} Depending on the model family, the fully automatic prediction is obtained from one of the session-level representations
\begin{equation}
\hat{y}_i^{\mathrm{text}} =
\begin{cases}
  f^{\mathrm{pool}}(\mathbf{z}_i^{\mathrm{pool}}) & \text{for Ridge and SVR} \\
  f^{\mathrm{mlp}}(\mathbf{z}_i^{\mathrm{mlp}}) & \text{for MLP} \\
  f^{\mathrm{seq}}(\mathbf{z}_i^{\mathrm{seq}}) & \text{for GRU and BiLSTM}
\end{cases}
\end{equation}

\noindent where $f^{\mathrm{pool}}$, $f^{\mathrm{mlp}}$, $f^{\mathrm{seq}}$ denote the family-specific regressors applied to the corresponding representations $\mathbf{z}_i$. 

\textbf{Interviewer integration.} The interviewer integration is combined with the raw interviewer rating using a fixed arithmetic mean:
\begin{equation}
    \hat{y}_i^{\mathrm{avg}} = \frac{\hat{y}_i^{\mathrm{text}} + y_i^{I}}{2}
\end{equation}
This combination requires no additional learned parameters and is fully transparent. The fused prediction is computed on held-out sessions after the fully automatic model has been trained on the corresponding training folds, so no information from the test fold is used in the combination.

\section{Evaluation}

\subsection{Training setup and evaluation metrics}

All evaluations are conducted on the 106 sessions that include complete patient and interviewer ratings, ensuring that the interviewer baseline, fully automatic models, and interviewer integration results are evaluated on a directly comparable set.

\begin{table*}[!t]
    \centering
    \caption{Hyperparameter search spaces used within the inner folds of nested GroupKFold.}
    \label{tab:hyperparameter}
    \begin{tabular}{ll}
        \toprule
        \textbf{Model} & \textbf{Searched hyperparameters} \\
        \midrule
        Ridge & $\alpha \in \{0.01, 0.1, 1, 10, 100\}$ \\
        SVR & $C \in \{0.1, 1, 10, 100\}$; $\gamma \in \{\texttt{scale}, 0.001, 0.01, 0.1\}$; $\epsilon \in \{0.01, 0.1, 0.5\}$ \\
        MLP & hidden dim $\in \{256, 512\}$; learning rate $\in \{10^{-4}, 3\times10^{-4}\}$; dropout $\in \{0, 0.1\}$; weight decay $\in \{0, 10^{-4}\}$ \\
        GRU and BiLSTM & recurrent hidden size $\in \{64, 128\}$; learning rate $\in \{10^{-4}, 3\times10^{-4}\}$; dropout $\in \{0, 0.1\}$; weight decay $\in \{0, 10^{-4}\}$ \\
        \bottomrule
    \end{tabular}
\end{table*}

\paragraph{Cross-validation protocol}
We adopt participant-level nested GroupKFold (5 outer folds and 4 inner folds) so that outer test folds contain only unseen participants. This is important because sessions from the same participant can share lexical habits, recurring topics, and reporting tendencies. Hyperparameter selection is confined to the inner training folds, and all reported predictions are obtained from held-out outer-fold evaluations.

\paragraph{Hyperparameter search}
Search spaces are summarized in Table~\ref{tab:hyperparameter}. The final search spaces were informed by preliminary development-stage runs and fixed before the reported nested-CV experiments. These pilot runs were used only to rule out clearly unsuitable configurations and to center the search around stable operating regions; all reported model selection and performance estimates were obtained within the nested cross-validation protocol. Neural models were trained using AdamW, early stopping, and up to 80 epochs.

\paragraph{Variance estimation}

To obtain stable performance estimates, we repeated the nested-CV procedure over 40 randomized runs for all models. For each run, we used a different seed to create the participant-level cross-validation splits. For neural models, we additionally varied the network initialization seed.

\paragraph{Metrics}
We report Pearson $r$, and mean absolute error (MAE). Pearson $r$ is the primary metric as it captures how well predictions recover the ordering of sessions, while MAE provides a complementary summary of absolute predictive error. They are widely used in prior work on automatic therapeutic alliance prediction~\cite{Goldberg2020, Lin2025}. We report the mean and standard deviation of held-out outer-fold metrics across the 40 randomized runs. For the main Pearson-correlation comparisons, we additionally report run-level bootstrap 95\% confidence intervals for the mean Pearson \(r\) and paired Wilcoxon signed-rank tests across runs.

\subsection{Overall Results}

Table~\ref{tab:overall} summarizes the main results on the overall patient-reported clinical interview experience prediction. The raw interviewer rating provided a competitive human-judgment baseline ($r=0.365$, MAE =15.717). Fully automatic models also achieved meaningful predictive performance, with the best fully automatic result obtained by the Ridge model ($r=0.286 \pm 0.044$). Among fully automatic models, SVR achieved the lowest MAE (15.125 $\pm$ 0.466), while the BiLSTM obtained \(r\) = 0.270 $\pm$ 0.084 and MAE = 15.256 $\pm$ 0.598.

\begin{table}[!t]
    \caption{Performance on the overall patient-reported clinical interview experience prediction. Values after $\pm$ indicate standard deviation across 40 runs.}
    \label{tab:overall}
    \centering

    \begin{tabular}{lll}
        \toprule
        \textbf{Prediction method} & \textbf{Pearson $r$} & \textbf{MAE} \\
        \midrule
        Interviewer & 0.365 & 15.717 \\
        \midrule
        \textit{Fully Automatic} \\
        \ \ \ Ridge & \textbf{0.286 $\pm$ 0.044}  & 16.717 $\pm$ 0.714 \\
        \ \ \ SVR & 0.240 $\pm$ 0.078  & \textbf{15.125 $\pm$ 0.466}  \\
        \ \ \ MLP & 0.197 $\pm$ 0.106 & 15.746 $\pm$ 0.773 \\
        \ \ \ GRU & 0.237 $\pm$ 0.111 & 15.405 $\pm$ 0.611 \\
        \ \ \ BiLSTM & 0.270 $\pm$ 0.084 & 15.256 $\pm$ 0.598 \\
        \midrule
        \textit{Interviewer Integration} \\
        \ \ \ Ridge & 0.390 $\pm$ 0.025 & 14.669 $\pm$ 0.359 \\ 
        \ \ \ SVR & 0.389 $\pm$ 0.023 & 14.310 $\pm$ 0.221 \\
        \ \ \ MLP & 0.376 $\pm$ 0.033 & 14.507 $\pm$ 0.339 \\
        \ \ \ GRU & 0.395 $\pm$ 0.036 & 14.295 $\pm$ 0.323 \\
        \ \ \ BiLSTM & \textbf{0.403 $\pm$ 0.030} & \textbf{14.214 $\pm$ 0.312} \\
        \bottomrule
    \end{tabular}
\end{table}

The strongest overall performance was obtained by the interviewer-integration setting, represented by the prediction $\hat{y}_i^{\mathrm{avg}}$. The BiLSTM interviewer-integration model achieved the highest Pearson correlation ($r = 0.403 \pm 0.030$; run-level bootstrap 95\% CI [0.393, 0.412]) and the lowest MAE ($14.214 \pm 0.312$). Across model types, the integrated prediction $\hat{y}_i^{\mathrm{avg}}$ consistently outperformed the raw interviewer baseline and improved over the corresponding fully automatic variants. Paired Wilcoxon signed-rank tests on Pearson $r$ confirmed the key comparisons: interviewer integration improved over interviewer-only by $\Delta r = +0.0378$, $p = 3.92 \times 10^{-9}$, and over fully automatic BiLSTM by $\Delta r = +0.1330$, $p = 9.09 \times 10^{-13}$. Overall, these results indicate that interviewer judgment and language-based prediction can be combined productively. To contextualize MAE, we evaluated a trivial training-mean baseline, which obtained MAE $=16.113 \pm 0.203$ across the 40 randomized runs. Compared with the patient-side composite SD of $18.9$, the MAE reduction of BiLSTM interviewer integration ($14.214 \pm 0.312$) is modest on the 0--100 scale, but improves upon both the mean predictor and interviewer-only baseline.

\begin{figure*}[!t]
    \centering
    \includegraphics[width=0.32\textwidth]{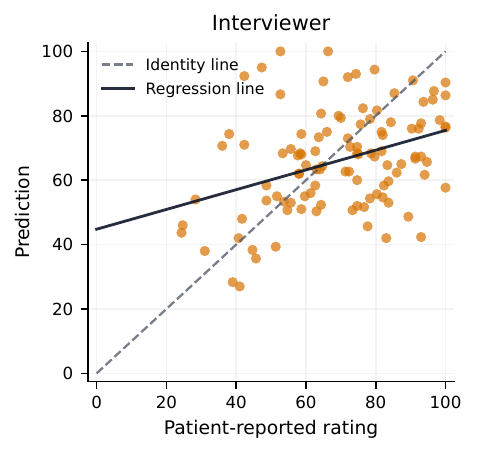}
    \hfill
    \includegraphics[width=0.32\textwidth]{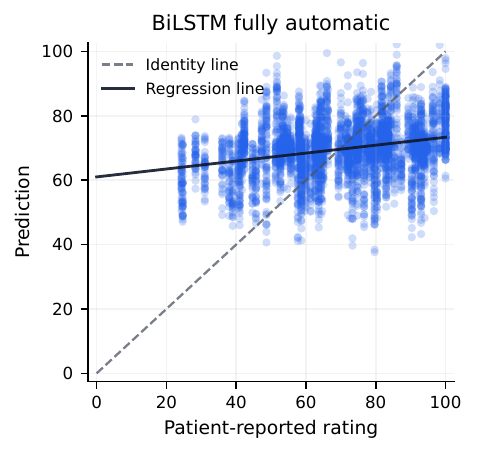}
    \hfill
    \includegraphics[width=0.32\textwidth]{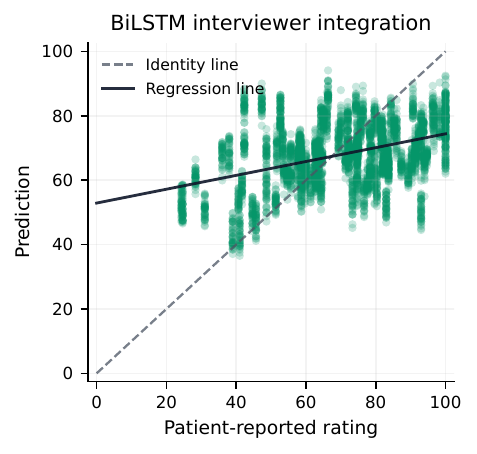}
    \caption{Session-level scatter plots for the matched held-out overall comparisons. Left: interviewer ratings. Middle: BiLSTM fully automatic prediction. Right: BiLSTM interviewer integration. The dashed line denotes the identity line, and the solid line denotes the fitted regression line.}
    \label{fig:scatter}
\end{figure*}

Figure~\ref{fig:scatter} shows scatter plots of predicted values versus patient-reported rating for interviewer-only predictions, BiLSTM fully automatic predictions, and BiLSTM interviewer integration. 
The interviewer panel (left) contains one point per matched session, whereas the BiLSTM and integration panels (middle and right) show held-out predictions from each of the 40 randomized runs. Therefore, each session appears 40 times in those two panels. 
The interviewer plot shows a moderate positive association with substantial scatter around the identity line, consistent with the competitive but imperfect correlation of $r=0.365$. The fitted line is shallower than the identity line, indicating that interviewer ratings tend to be compressed toward the middle of the scale.
The BiLSTM fully automatic plot (middle) also shows a positive trend ($r=0.270 \pm 0.084$), but its predictions occupy an even narrower vertical range than the interviewer ratings. The corresponding regression line is therefore shallower, indicating stronger range compression and an underestimation of session-to-session variability in patient-reported ratings. The BiLSTM interviewer integration plot (right) shows improved agreement overall ($r=0.403 \pm 0.030$), with predictions shifted closer to the identity line than in the fully automatic case. However, the fitted regression line remains shallower than the identity line, so the integrated prediction still exhibits range compression even though its overall correlation and error are improved.

\subsection{Speaker Ablation}

To analyze where the predictive signal resides within the conversation, we evaluate three input configurations for the language-based models: \textit{patient-only}, which uses only the patient embedding stream; \textit{interviewer-only}, which uses only the interviewer embedding stream; and \textit{dual-stream}, which combines both streams by late fusion. Table~\ref{tab:ablation} compares patient-only, interviewer-only, and dual-stream inputs in the fully automatic setting for Ridge and BiLSTM, which were the best-performing pooled and sequence models in Table~\ref{tab:overall}. 

Across both model types, patient-only models were the weakest, while interviewer-only inputs achieved the strongest Pearson correlations. For Ridge, interviewer-only reached $r=0.298 \pm 0.040$, compared with $r=0.286 \pm 0.044$ for dual-stream and $r=0.192 \pm 0.069$ for patient-only. For BiLSTM, interviewer-only reached $r=0.285 \pm 0.106$, compared with $r=0.270 \pm 0.084$ for dual-stream and $r=0.079 \pm 0.119$ for patient-only. MAE showed a similar pattern for Ridge, while the BiLSTM dual-stream model achieved slightly lower MAE than interviewer-only. Overall, these results suggest that interviewer-side language carries the most readily extractable predictive signal in the fully automatic setting, whereas patient-side language alone is less predictive under the current modeling setup.

\begin{table}[!t]
    \caption{Comparison of patient-only, interviewer-only, and dual-stream inputs for Ridge and BiLSTM in the fully automatic setting. Values after $\pm$ indicate standard deviation across 40 runs.}
    \label{tab:ablation}
    \centering

    \begin{tabular}{lll}
        \toprule
        \textbf{Input streams} & \textbf{Pearson $r$} & \textbf{MAE} \\
        \midrule
        \textit{Ridge} \\
        \ \ \ Dual stream & 0.286 $\pm$ 0.044 & 16.717 $\pm$ 0.714 \\
        \ \ \ Interviewer-only & \textbf{0.298 $\pm$ 0.040} & \textbf{16.519 $\pm$ 0.830}\\
        \ \ \ Patient-only & 0.192 $\pm$ 0.069 & 17.098 $\pm$ 0.935 \\
        \midrule
        \textit{BiLSTM} \\
        \ \ \ Dual stream & 0.270 $\pm$ 0.084 & \textbf{15.256 $\pm$ 0.598} \\
        \ \ \ Interviewer-only & \textbf{0.285 $\pm$ 0.106} & 15.323 $\pm$ 0.860 \\
        \ \ \ Patient-only & 0.079 $\pm$ 0.119 & 16.490 $\pm$ 1.003 \\
        \bottomrule
    \end{tabular}
\end{table}

\subsection{Pooling Strategy Ablation}

For the BiLSTM model, we additionally compare three pooling strategies: masked mean pooling, attention pooling, and joint attention pooling. Table~\ref{tab:pooling} shows that masked mean pooling consistently matched or outperformed both attention pooling and joint attention pooling in both the fully automatic and the interviewer integration setting. As the simpler and slightly better performing strategy, masked mean pooling is therefore used in all other reported BiLSTM results.

\begin{table}[!t]
    \caption{Ablation of pooling using BiLSTM. Values after $\pm$ indicate standard deviation across 40 runs.}
    \label{tab:pooling}
    \centering

    \begin{tabular}{lll}
        \toprule
        \textbf{Pooling} & \textbf{Pearson $r$} & \textbf{MAE} \\
        \midrule
        \textit{Fully Automatic} \\
        \ \ \ Masked mean pooling & \textbf{0.270 $\pm$ 0.084}& \textbf{15.256 $\pm$ 0.598} \\
        \ \ \ Attention pooling & $0.254 \pm 0.099$ & $15.490 \pm 0.728$ \\
         \ \ \ Joint attention & $0.185 \pm 0.116$ & $15.916 \pm 0.832$ \\
        \midrule
        \textit{Interviewer Integration} \\
        \ \ \ Masked mean pooling & \textbf{0.403 $\pm$ 0.030} & 
        \textbf{14.214 $\pm$ 0.312} \\
        \ \ \ Attention pooling & $0.399 \pm 0.034$ & $14.239 \pm 0.373$ \\
         \ \ \ Joint attention & $0.373 \pm 0.040$ & $14.555 \pm 0.414$ \\
        \bottomrule
    \end{tabular}
\end{table}

\section{Discussion and Limitations}

\textbf{Main findings.} Our results support three main conclusions. First, patient-reported clinical interview experience can be estimated from the language used during the interaction, as fully automatic models achieved meaningful predictive performance on the overall target. Second, interviewer ratings provided a competitive human-judgment baseline, but the strongest overall results were obtained when interviewer ratings were combined with language-based predictions. Third, within the fully automatic setting, interviewer-side language was more informative than patient-side language in terms of Pearson correlation, while dual-stream inputs did not consistently outperform interviewer-only inputs.

The consistent benefit of integration across all five model types is the central finding of this study. It demonstrates that the benefit is not an artifact of a particular architectural choice, but reflects a genuine
complementarity between the two information sources. Interviewer ratings reflect a post-session, in-context human evaluation, whereas automatic predictions derive their estimates from distributional patterns in the transcript. The performance gains from combining these sources therefore indicate that they capture different, rather than redundant, aspects of the patient's experience.

The speaker-stream ablation offers a more fine-grained account of where this signal originates. The finding that interviewer-side language is more informative than patient-side language suggests that supportive session experience is partly constituted by interviewer behaviour, including how the interviewer structures the session, responds to disclosures, and creates space for the patient to speak. Patient language, by contrast, may be more
strongly shaped by symptom severity, idiosyncratic disclosure style, and topic selection, making it a noisier signal with respect to patient-reported experience. Dual-stream inputs remained competitive, but did not consistently improve over interviewer-only inputs under the present modeling setup. 
The two streams are therefore asymmetric in informativeness: interviewer-side language appears to contain the strongest readily extractable signal for this task, while patient-side language alone is less predictive.

These findings extend previous psychotherapy research that establishes that patient and clinician perspectives are related but not interchangeable~\cite{hartmann2015accuracy, atzil2019therapists} to the computational domain, showing that automatic language-based predictions carry signal partly independent of interviewer evaluation. Rather than replacing human judgment, automatic prediction appears most useful as an additional evidence source, particularly when the goal is to approximate the patient's perspective. The simplicity of the integration strategy, a fixed arithmetic
average requiring no additional training, makes this approach practically deployable.

\textbf{Limitations.} Several limitations should be noted.
First, the dataset is small (107 sessions, 106 with interviewer ratings), limiting the stability of model comparisons and the generalisability of conclusions. Replication across larger, more diverse corpora covering different clinical settings, patient populations, and languages is necessary before broader claims can be made.
Secondly, while these three questionnaire items captured complementary aspects of patient-reported interaction quality, they should not be interpreted as a complete operationalization of the therapeutic alliance, nor should our findings be directly equated with results obtained in an ongoing psychotherapy setting.
Third, we deliberately focused on text-based modelling for reasons of
privacy, scalability, and deployment feasibility; the present results therefore characterise what can be recovered from transcript-based representations rather than establishing an upper bound on multimodal approaches.
Fourth, the arithmetic mean fusion may underestimate the gains achievable with learned combination strategies, which future work with larger samples should explore. Finally, the finding that interviewer-side language was more predictive than patient-side language should not be interpreted causally; it reflects a more readily extractable signal under the current data and modelling setup rather than evidence that interviewer behaviour alone determines patient experience.

\section{Conclusion}
We investigated whether language-based automated predictions could combine patient-reported interaction quality predictions with post-interview interviewer ratings in free-flowing clinical interviews. We evaluated this integration approach across five standard models using 107 interview transcripts from the MePheSTO corpus. The results showed that a simple average of automated predictions and interviewer ratings consistently outperformed either source alone. This consistent advantage suggests that automatic language analysis and interviewer judgment provide complementary predictive information.

The simplicity and consistency of our integration strategy suggest that it can be used in a deployable framework in which automated language analysis tools can enhance, rather than replace, post-clinical assessments.

\section*{Ethical Impact Statement}

This study analyzes transcript-based representations of psychiatric interviews from the German MePheSTO corpus to estimate patients' subjective experience of a clinical conversation. The study was approved by the medical ethics committee of the Carl-von-Ossietzky University of Oldenburg (approval number 2021-108) and conducted in accordance with the latest revision of the Declaration of Helsinki. Participants provided written informed consent after receiving a complete description of the study. The data are confidential due to the sensitivity of psychiatric information and the potential risks of privacy breaches. Accordingly, this work is limited to research use on de-identified conversational transcripts and does not involve automated clinical decision-making or deployment in patient care. We stress that the models evaluated here are intended to support research on interaction quality and should not replace clinician judgment or patients' own reports. Any future clinical application would require further validation, clear governance and safeguards against harmful or inappropriate use. The generalizability of our findings may be limited, as we only studied patients from a single clinic in Germany who had a Major Depressive Episode.

\section*{Acknowledgment}
This research was in parts funded by the French National Research Agency ANR under the UCA\textsuperscript{JEDI} Investments into the Future (project number ANR-15-IDEX-01) and by the German Ministry for Education and Research BMBF (grant number 01IS20075).

\bibliographystyle{IEEEtran}
\bibliography{references}

\end{document}